\magnification=\magstephalf
\hsize=16 truecm
\vsize=24 truecm
\hoffset=-0.1truecm

\parindent=12 pt

\footline={\ifnum\pageno>0 \tenrm \hfil\folio\hfil \else \hfil\fi}

%
%
\font\tenbi=cmmib10 
\skewchar\tenbi='177 
\font\eightbi=cmmib7 at 8pt 
\font\sevenbi=cmmib7 
\skewchar\sevenbi='177 
\font\sixbi=cmmib5 at 6pt 
\font\fivebi=cmmib5 
\skewchar\fivebi='177 
\newfam\bmitfam \def\bi{\fam\bmitfam\tenbi} 
\textfont\bmitfam=\tenbi 
\scriptfont\bmitfam=\sevenbi 
\scriptscriptfont\bmitfam=\fivebi 

\mathchardef\alpha="710B 
\mathchardef\beta="710C 
\mathchardef\gamma="710D 
\mathchardef\delta="710E 
\mathchardef\epsilon="710F 
\mathchardef\zeta="7110 
\mathchardef\eta="7111 
\mathchardef\theta="7112 
\mathchardef\iota="7113
\mathchardef\kappa="7114 
\mathchardef\lambda="7115 
\mathchardef\mu="7116 
\mathchardef\nu="7117 
\mathchardef\xi="7118 
\mathchardef\pi="7119 
\mathchardef\rho="711A 
\mathchardef\sigma="711B 
\mathchardef\tau="711C 
\mathchardef\upsilon="711D 
\mathchardef\phi="711E 
\mathchardef\chi="711F 
\mathchardef\psi="7120 
\mathchardef\omega="7121 
\mathchardef\varepsilon="7122 
\mathchardef\vartheta="7123 
\mathchardef\varpi="7124 
\mathchardef\varrho="7125 
\mathchardef\varsigma="7126 
\mathchardef\varphi="7127 
 
\def\epsjlon{\varepsilon}

\def\ki{{\raise1pt\hbox{$\chi$}}}

\def\mbi#1{{\bi #1}}

\font\eighti=cmmi8 
\font\sixi=cmmi6
\font\sixrm=cmr6
\font\eightrm=cmr8 
\font\ninerm=cmr9
\font\eightit=cmti8
\font\sixbf=cmbx6 
\font\eightbf=cmbx8 
\font\eightsy=cmsy8 
\font\sixsy=cmsy6 

\font\teneusm=eusm10
\def\fl#1{\hbox{\teneusm#1}}
\font\sc cmcsc10
\font\abs cmbx9

%
%
\def\eightpoint{\def\rm{\fam0\eightrm}
  \textfont0=\eightrm \scriptfont0=\sixrm \scriptscriptfont0=\fiverm
  \textfont1=\eighti  \scriptfont1=\sixi  \scriptscriptfont1=\fivei
  \textfont2=\eightsy \scriptfont2=\sixsy \scriptscriptfont1=\fivesy
  \textfont3=\tenex   \scriptfont3=\tenex \scriptscriptfont3=\tenex
  \textfont\itfam=\eightit  \def\it{\fam\itfam\eightit}
  \textfont\bffam=\eightbf  \scriptfont\bffam=\sixbf
   \scriptscriptfont\bffam=\fivebf  \def\bf{\fam\bffam\eightbf}%
  \textfont\bmitfam=\eightbi  \scriptfont\bffam=\sixbi
   \scriptscriptfont\bmitfam=\fivebi  \def\bi{\fam\bmitfam\eightbi}%
  \normalbaselineskip=9pt
  \setbox\strutbox=\hbox{\vrule height7pt depth2pt width0pt}%
  \let\sc=\eightsc  \let\big=\eightbig \normalbaselines\rm}

\def\eightbig#1{{\hbox{$\textfont0=\ninerm\textfont2=\ninesy
  \left#1\vbox to6.5pt{}\right.\n@space$}}}

%
%
\def\tenpoint{\def\rm{\fam0\tenrm}
  \textfont0=\tenrm \scriptfont0=\sevenrm \scriptscriptfont0=\fiverm
  \textfont1=\teni  \scriptfont1=\seveni  \scriptscriptfont1=\fivei
  \textfont2=\tensy \scriptfont2=\sevensy \scriptscriptfont1=\fivesy
  \textfont3=\tenex   \scriptfont3=\tenex \scriptscriptfont3=\tenex
  \textfont\itfam=\tenit  \def\it{\fam\itfam\tenit}
  \textfont\bffam=\tenbf  \scriptfont\bffam=\sevenbf
   \scriptscriptfont\bffam=\fivebf  \def\bf{\fam\bffam\tenbf}%
  \textfont\bmitfam=\tenbi  \scriptfont\bffam=\sevenbi
   \scriptscriptfont\bmitfam=\fivebi  \def\bi{\fam\bmitfam\tenbi}%
  \normalbaselineskip=9pt
  \setbox\strutbox=\hbox{\vrule height8.5pt depth3.5pt width0pt}%
  \let\sc=\sc  \let\big=\tenbig \normalbaselines\rm}

\def\tenbig#1{{\hbox{$\left#1\vbox to8.5pt{}\right.\n@space$}}}

\input amssym.def \input amssym.tex 

\def\hs#1{\hskip #1pt}
\def\vs#1{\vskip #1pt}

\def\displayskip{\noalign{\vskip 9pt plus 3pt minus 4pt}}
\def\smalldisplayskip{\noalign{\vskip 5pt plus 1pt minus 2pt}}

\def\ts{\textstyle}
\def\ds{\displaystyle}
\def\ss{\scriptstyle}
\def\sss{\scriptscriptstyle}

\def\diff{{\rm d}}
\def\ket#1{\vert #1 \rangle}
\def\bra#1{\langle #1 \vert}
\def\norm#1{\Vert #1 \Vert}

\def\llongrightarrow{\relbar\mkern-10mu{\relbar\mkern-10mu
{\relbar\mkern-10mu{\longrightarrow}}}}

\def\buildler#1\over#2{\mathrel{\mathop{#1}\limits_{#2}}}
\def\ttofor#1{\buildler \llongrightarrow \over {#1}}

\def\si#1#2{{\vrule height#1pt depth#2pt width0pt}}
\def\compint{\!\! \int \!\!}
\def\ncdot{\! \cdot \!}

\def\fracv{\raise 2pt\hbox{,}}
\def\fracp{\raise 2pt\hbox{.}}

%
%
\global\newcount\noteno \global\noteno=1
\newwrite\notefile

\def\note#1#2{\xdef#1{\the\noteno}%
\ifnum\noteno=1\immediate\openout\notefile=notes\fi%
\immediate\write\notefile{\noexpand\item{\noexpand#1.\ }#2}%
\global\advance\noteno by1}

\def\scriptnote#1{$^{{\rm #1}}$}

\def\pg#1{\hbox{p.\hs{2}#1}} 

\input epsf.tex

\note\bellsntnc{
J. S. Bell, 
in {\it Schr\"odinger. Centenary celebration of a polymath\/}, 
C. W. Kilmister, ed. (Cambridge University Press, Cambridge, 1987), pg{41}.
}
\note\gottfried{
K. Gottfried, {\it Quantum Mechanics\/} (Benjamin, New York, 1966). 
}
\note\nrXCI{
O. Nicrosini and A. Rimini, 
in {\it Symposium on the Foundations of Modern Physics 1990\/}, 
P. Lahti and P. Mittelstaedt, eds. (World Scientific, Singapore, 1991), \pg{280}.
}
\note\riminiXCIV{
A. Rimini, 
in {\it Proceedings of the Cornelius Lanczos International Centenary Conference\/}, 
J. D. Brown, M. T. Chu, D. C. Ellison, and R. J. Plemmons, eds., 
(SIAM, Philadelphia, 1994), pg{591}. 
}
\note\bohmLII{
D. Bohm, Phys. Rev. {\bf 85} (1952), 166, 180.
}
\note\griffithsLXXXIV{
R. Griffiths, J. Stat. Phys. {\bf 36} (1984), 219.
}
\note\gmhXCIII{
M. Gell-Mann and J. Hartle, Phys. Rev. D {\bf 47} (1993), 3345.
}
\note\omnesXCII{
R. Omn\`es, Rev. Mod. Phys. {\bf 64} (1992), 339.
}
\note\grw{
G. C. Ghirardi, A. Rimini, and T. Weber, 
Phys.\ Rev.\ D {\bf 34}, 470 (1986); {\bf 36} 3287 (1987). 
}
\note\pearleLXXXIX{
P. Pearle, 
Phys. Rev. A {\bf 39} (1989), 2277.
}
\note\gprXC{
G.C. Ghirardi, P. Pearle, and A. Rimini, 
Phys. Rev. A {\bf 42} (1990), 78.
}
\note\psXCIV{
P. Pearle and E. Squires, 
Phys.\ Rev.\ Lett.\ {\bf 73} (1994), 1. 
}
\note\bnr{
M. Buffa, O. Nicrosini, and A. Rimini, 
Found.\ Phys.\ Lett.\ {\bf 8} (1995), 105. 
}
\note\pearleXC{
P. Pearle, 
in {\it Sixty--Two Years of Uncertainty\/}, A. Miller, ed., (Plenum, New York, 
1990), pg{193}.
}
\note\ggpXC{
G. C. Ghirardi, R. Grassi and P. Pearle, 
Foundations of Physics {\bf 20} (1990), 1271. 
}
\note\pearleIC{
P. Pearle, 
Physical Review A {\bf 59} (1999), 80.
}

\newcount\numsez
\newcount\numeq
\numsez=0
\numeq=0
\def\sez#1{\global\numeq=0\global\advance\numsez by 1\vs{18}%
\leftline{\bf\the\numsez. #1}\vs9}
\def\numera{\global\advance\numeq by 1\leqno(\the\numsez.\the\numeq)}

\def\pr#1{\lower#1pt\hbox{$'$}}

\def\sInt{s\raise2.5pt\hbox{\sevenrm I}}
\def\sCl{s\raise2.5pt\hbox{\sevenrm C}}
\def\Tmumu{T\si{6.5}{0}^\mu\raise1.5pt\hbox{$\si{1}{0}_{\hs{-1.5}\mu}$}(\bar x)}

\def\domtext{$D\lower2pt\hbox{\hs{-1}$\ss t$ \kern -5.5pt
\vrule width 2.2pt height 5.3pt depth -5.1pt}(t,\mbi x)$}

\def\domscript{{\!\ss D\lower1.6pt\hbox{\hs{-1}$\sss t$ \kern-5.25pt\vrule 
width 1.6pt height 3.90pt depth -3.70pt}(t,\mbi x)}}

\def\centercol#1#2{\centerline{\vtop{\hsize #1truecm #2}}}

\pageno=0
\rightline{FNT/T 2002/11}
 
\vskip 4.5truecm 
\centerline{\bf RELATIVISTIC SPONTANEOUS LOCALIZATION:}%
\bigskip
\centerline{\bf A PROPOSAL}
\vskip 1truecm 

\centerline{{\sc Oreste Nicrosini}\footnote{$^1$}{\eightrm E--mail: 
nicrosini@pv.infn.it}}
\vskip 6pt
\centerline{\it Istituto Nazionale di Fisica Nucleare, Sezione di Pavia, Pavia}
\vskip 0.5truecm
\centerline{{\sc Alberto Rimini}\footnote{$^2$}{\eightrm E--mail: 
rimini@pv.infn.it}}
\vskip 6pt
\centerline{\it Universit\`a di Pavia, Dipartimento di Fisica Nucleare 
e Teorica, Pavia}
\centerline{\it Istituto Nazionale di Fisica Nucleare, Sezione di Pavia, Pavia}

\vskip2truecm

\centercol{10}{\noindent{\abs Abstract ---} {\ninerm 
A new proposal for a Lorentz--invariant spontaneous localization theory is presented. It 
is based on the choice of a suitable set of macroscopic quantities to be stochastically 
induced to have definite values. Such macroscopic quantities have the meaning of a 
time--integrated amount of a microscopically defined quantity called stuff related to the 
presence of massive particles. 
}}

\vfill\eject 

\sez{\bf Introduction}

{\it Either the wave function, as given by the Schr\"odinger equation is not everything, 
or it is not right.\/} John Bell\scriptnote{\bellsntnc} expressed with this cathegoric 
sentence his point of view about the problem of quantum measurement. A possible way out 
from this dylemma consists in assuming that the wave function describes statistical 
ensembles rather than individual systems and resorting to the practical impossibility 
(decoherence) of detecting interference between the different macroscopically 
distinguishable terms appearing in the wave function after a measurement. The latter 
interpretative attitude\scriptnote{\gottfried} works to a considerable extent, but it 
cannot avoid certain typical inconsistencies\scriptnote{\nrXCI,\riminiXCIV} which arise 
from forgoing any tool apt to describe the result of an individual measurement. If 
the outcome of an individual measurement is to have a counterpart in the description of 
the system after the measurement, then one is led back to Bell's dylemma. 

That the wave function is not everything means that there are additional variables which, 
together with the wave function, constitute the state of the system. A theory of this 
kind is Bohm's pilot--wave formulation of quantum mechanics,\scriptnote{\bohmLII} where 
the additional variables are identified with the configuration of the system. 
Also the branch labels used to describe the system in the history approach to quantum 
mechanics\scriptnote{\griffithsLXXXIV-\omnesXCII} can be considered as additional 
variables. 

That the wave function is not right means that a modification of the Schr\"odinger 
equation is to be accepted. As a matter of fact, the reduction principle of the standard 
formulation itself is a modification of the Schr\"odinger equation, a modification which 
allows to interpret the wave function as describing an individual system. The reduction 
principle, however, can be formulated only accompanied by the ambiguous distinction 
between quantum systems and measuring apparatus. Reduction theories, instead, describe 
reduction by a definite, mathematically precise correction to the Schr\"odinger equation, 
the corrected equation being supposed to be valid in any circumstance. The correction 
must rapidly reduce superpositions of macroscopically distinguishable states and, 
nevertheless, have practically unobservable consequences in all ordinary situations. 
Reduction theories can be considered as quantitative versions of the standard reduction 
principle. They are necessarily stochastic and nonlinear, just because the reduction 
principle is such. 

In reduction theories\scriptnote{\grw-\bnr} a stochastic process is introduced which 
induces the quantities belonging to a suitable set to have definite values. Such 
quantities are defined quantum mechanically but have a macroscopic character. They are 
always related to position, so that the result of the process is a localization of 
macroscopic objects. The localization is spontaneous in the sense that we postulate its 
existence at the level of the fundamental equation of quantum mechanics, without 
attempting to find its origin in terms of a level of description going beyond that. 
Successful reduction theories are all based on a spontaneous localization process which 
becomes effective only when the macroscopic level is reached. The reduction of the wave 
function of a measured microscopic quantum mechanical system takes place via the 
micro--macro correlations settled by the measuring device. 

As it will be described in sect.\ 3, the definition of the macroscopic quantities 
involves an integration over a tiny space region, of a spherical shape for reasons of 
invariance of the theory. Such spherical regions cannot reduce to a point, because the 
quantities would lose their macroscopic character, exhibiting fluctuations related to 
the microscopic structure of the system. Furthermore, reducing to a point such regions 
would be equivalent to a pointlike localization of the constituent particles with obvious 
disastrous consequences. 

The theories sketched above are nonrelativistic. The difficulty met in constructing a 
relativistic generalization is that the spherical regions mentioned above are not Lorentz 
invariant. Nevertheless, relativistic spontaneous localization models have been proposed. 
In the first model\scriptnote{\pearleXC,\ggpXC} a pointlike process acts on a light boson 
field which in turn is coupled to the fields describing the common material particles. In 
such a way the material particles are localized within regions whose dimensions are ruled 
by the light mass of the boson field. A pointlike process is nevertheless there, and this 
causes, e.g., an infinite rate of energy production. A second and a third model%
\scriptnote{\pearleIC} cure this problem partially and completely, respectively. We find 
it somewhat difficult, however, to grasp the physical meaning of the assumptions 
underlying the models. 

In our opinion, compelling certain macroscopic quantities to have definite values is the 
essential feature of reduction theories. We shall try in the following to found the 
construction of a relativistic spontaneous localization model on the identification of a 
suitable set of macroscopic quantities to be induced by a stochastic process to have 
definite values. 

In sect.\ 2 we describe the class of Markov processes in Hilbert space which induce the 
state vector to move towards the eigenspaces of the operator representing a quantity (or 
the common eigenspaces of the operators representing a set of compatible quantities). In 
sect.\ 3 the most reliable nonrelativistic reduction theory is concisely presented. In 
sects.\ 4 and 5 the framework for a relativistic reduction model is introduced. In 
sects.\ 6 and 7 we put forward a proposal for the quantities to be induced to have 
definite values and discuss it. In the final section we list some open problems and 
conclude. 
 
\sez{\bf Markov processes in Hilbert space}

Given a Hamiltonian operator $H$ and a selfadjoint linear operator $A$, let us consider 
the (It\^o) stochastic differential equation 
$$
\diff\ket{\psi(t)} = \left[-{i \over \hslash}\,H\,\diff t + g\,A_{\psi(t)}\diff B(t) 
- \ts{1\over2}\, g^2 \!\left(A_{\psi(t)}\right)^2\!\diff t\right] \ket{\psi(t)} , 
\numera
$$
where $A_\psi$ is the nonlinear operator 
$$
A_{\psi} = A - \bra{\psi} A \ket{\psi} 
\numera
$$
and $B(t)$ is a Wiener process such that 
$$
\overline{\diff B(t)} = 0, ~~~~~~~~~~\overline{(\diff B(t))^2} = \diff t . 
\numera
$$
Eq.\ (2.1) conserves the norm of the state vector $\ket\psi$. 

One can prove\scriptnote{\gprXC} that, if the Schr\"odinger term is dropped, the 
solutions of the resulting equation 
$$
\diff\ket{\psi(t)} = 
\left[g\,A_{\psi(t)}\diff B(t)
- \ts{1\over2}\, g^2 \!\left(A_{\psi(t)}\right)^2\!\diff t \right] \ket{\psi(t)} 
\numera
$$
have the limit 
$$
\ket{\psi(t)} ~~\ttofor{t \rightarrow \infty}~~ 
P_{\!\epsjlon} \ket{\psi(t_0)} \big/ \norm{P_{\!\epsjlon} {\psi(t_0)}} , ~~~~~~~~
\Pr(\epsjlon) = \norm{P_{\!\epsjlon} {\psi(t_0)}}^2,
\numera
$$
where $P_{\!\epsjlon}$ are the projection operators on the eigenspaces of $A$ (with 
obvious modifications in the case of continuous spectrum). It is seen that the stochastic 
term drives $\ket\psi$ towards an eigenstate of $A$; or, in other words, it compels in 
the long run the quantity described by $A$ to have a definite value. The process becomes 
ineffective when $\ket{\psi(t)}$ reaches an eigenvector of $A$ because $A_\psi$ is zero 
when applied to such a vector. At any time $t_0$, the probability of ending up in a 
definite eigenspace is the square norm of the projection of the state vector at time 
$t_0$ on that eigenspace and, in the case of degeneracy, the precise final eigenvector is 
given by the projection rule. 

If both the Schr\"odinger term and the stochastic term are kept, the net result will 
depend on the competition, if it is there, between the two evolution processes. 

If, instead of a single $A$, a set of commuting selfadjoint operators $A^i$ is 
considered, the single stochastic term built with the operator $A$ is replaced by one 
term for each $A^i$ and the stochastic evolution equation becomes 
$$
\diff\ket{\psi(t)} = 
\left[-{i \over \hslash}\,H\,\diff t + 
{\sum}_i \left( g_i\,A^i_{\psi(t)}\,\diff B_i(t) - {\ts{1\over2}}\, 
g_i^2 \left(\!A^i_{\psi(t)}\!\right)^2\!\diff t \right)\right] \ket{\psi(t)} , 
\numera
$$
where 
$$
\overline{\diff B_i(t)} = 0, ~~~~~~~~~~
\overline{\diff B_i(t)\diff B_{j}(t)} = \delta_{ij}\diff t . 
\numera
$$
If the Schr\"odinger term is dropped, the limit of the solutions is still given by 
eq.\ (2.5) where $P_{\!\epsjlon}$ are now the projection operators on the common 
eigenspaces of the operators $A^i$.

\sez{The nonrelativistic mass process} 

The quantities being driven by the stochastic process to have definite values are the 
mass densities $D(\mbi x)$ averaged over small macroscopic spherical domains around all 
space points $\mbi x$\scriptnote{\bnr}. The evolution equation for the state vector 
$\ket{\psi(t)}$ is then 
$$
\diff \ket{\psi(t)} = \bigg[-{i \over \hslash}\, H\, \diff t 
+ \compint \diff^3 \hs{-.6}\mbi x \,\bigg( \!{g_0 \over \raise2pt\hbox{$m_0$}}
\,D_{\!\psi(t)}\!(\mbi x) \,\diff B_{\mbi x} (t)- {\ts{1\over2}} 
\left(\!{g_0 \over \raise2pt\hbox{$m_0$}}\!\right)^{\!2}\!
\left(\! D_{\!\psi(t)}\!(\mbi x)\!\right)^{\!2} \!\diff t\!\bigg)\bigg]\ket{\psi(t)} , 
\numera 
$$
where 
$$
D_{\!\psi(t)}\!(\mbi x) = D(\mbi x) - \bra{\psi(t)} D(\mbi x) \ket{\psi(t)} 
\numera
$$
and the stochastic field $B_{\mbi x}(t)$ has the properties 
$$
\overline{\diff B_{\mbi x}(t)}=0 , ~~~~~~ 
\overline{\diff B_{\mbi x}(t)\,\diff B_{\mbi x'}(t)} = 
\delta^{\sss(\hs{-.5}3\hs{-.5})}\hs{-.5}(\mbi x - \mbi x\pr{1.5}) \,\diff t . 
\numera
$$
The strength constant $g_0$ is defined with respect to a reference mass $m_0$. 

The macroscopic mass densities $D(\mbi x)$ are defined by 
$$
D(\mbi x) = \compint \diff^3 \hs{-.6}\bar\mbi x \,
F(\bar\mbi x - \mbi x) \,m(\bar\mbi x)
\numera
$$
in terms of the microscopic mass density $m(\mbi x)$. The function $F(\mbi x)$ identifies 
the spherical domain over which the mass densities are averaged. It can be chosen to be 
the smooth function 
$$
\hbox to 95pt{}
F(\mbi x) = {\cal N}
\exp\left( -{\ts{1\over2}} \big({\mbi x/a}\big)^{\!2} \right) , 
\hbox to 55pt{}
{\cal N} = \Big(\!{\lower2pt\hbox{$1$} \over 
\,2\hbox{$\pi a^2$}}\!\Big)^{\!\!\sss {3\over2}} , 
\numera
$$
where the length constant $a$ defines the linear dimensions of the small macroscopic 
domain. The microscopic mass density is 
$$
m(\mbi x) = \,{\sum}_k m_k {\sum}_s
\,a^{\sss\!+}_k(\mbi x,s) \,a_k(\mbi x,s)
\numera
$$
where $k$ runs over the different kinds of identical particles and, for each $k$, $m_k$ 
is the mass of the particles, $a^{\sss\!+}_k(\mbi x,s)$, $a_k(\mbi x,s)$ are the creation 
and annihilation operators of particles at $\mbi x$ with spin component $s$. 

We remark that shrinking to a point the spherical domain which defines densities (i.e.\ 
using $m(\mbi x)$ instead of $D(\mbi x)$) is not permitted, not just because the 
consequences are unacceptable, but, more deeply, because the quantities driven to have a 
definite value must have a macroscopic meaning. On the other hand, the spherical domain 
can neither be chosen too large, because the set of densities $D(\mbi x)$ would 
lose its power of distinguishing macroscopically distinguishable situations. 

The suggested values of the parameters $a$ and $g_0$ are 
$$
a \approx 10^{-5} \,{\rm cm}, ~~~~~~~~~ 
g_0^2 \approx 10^{-30} \,{\rm cm^3 \, s^{-1}} , 
\numera 
$$
having taken the proton mass as the reference mass $m_0$. It has been shown%
\scriptnote{\gprXC-\bnr} that with such a choice superpositions of macroscopically 
distinguishable states are quickly reduced, while for the rest one has only negligible 
effects. Obviously, the theory is not exactly equivalent to standard quantum mechanics 
and is in principle falsifiable. 

The precise form of the function $F(\mbi x)$ is practically unimportant. An alternative 
choice is the square function 
$$
\hbox to 107pt{}
F(\mbi x) = {\cal N}\,
\ki_{(0,a^2)}\left(\mbi x^2\right) , 
\hbox to 80pt{}
{\cal N} = {\lower2pt\hbox{$3$} \over \,4\hbox{$\pi a^3$}} \,\fracv
\numera
$$
where $\ki_E(x)$ is the characteristic function of the set $E$. 

Instead of using the mass densities averaged over the considered spherical volumes, one 
can equivalently express the process in terms of the total masses contained in the same 
volumes. Then the evolution equation becomes 
$$\eqalign{
\diff \ket{\psi(t)} = \bigg[&-{i \over \hslash}\, H\, \diff t \cr 
&+ \compint \diff^3 \hs{-.7}\mbi x \, \bigg( \!g
\left(\!M_{\psi(t)}\!(\mbi x)\big/m_0\!\right) \diff B_{\mbi x} (t)
- {\ts{1\over2}}\, g^2
\!\left(\!M_{\psi(t)}\!(\mbi x)\big/m_0\!\right)^{\!2} \!\diff t\!
\bigg)\bigg]\ket{\psi(t)} , 
}
\numera
$$
where 
$$\eqalign{
&M_{\psi(t)}\!(\mbi x) = M(\mbi x) - \bra{\psi(t)} M(\mbi x) \ket{\psi(t)} , 
\cr \smalldisplayskip
&M(\mbi x) = \compint \diff^3 \hs{-.6}\bar\mbi x 
\,\ki_{(0,a^2)}\left((\bar\mbi x-\mbi x)^2\right) \,m(\bar\mbi x) \, . 
}
\numera
$$
The squared strength constant $g^2$ is 
$$
g^2 = \left(\!{\lower2pt\hbox{$3$} \over \,4\hbox{$\pi a^3$}}\!\right)^{\!2} g_0^2 
\approx 6\,\,10^{-2} \,{\rm cm^{-3} \, s^{-1}} . 
\numera
$$

\sez{\bf The Tomonaga--Schwinger equation}

In order to write down a Lorentz--invariant version of eq.\ (3.9), it is convenient to 
start from a manifestly covariant form of the Schr\"odinger equation. This is provided by 
the Tomonaga--Schwinger interaction--picture (IP) equation. 

To simplify the notation from now on we shall use $\hslash=c=1$ units. 

In the Tomonaga--Schwinger approach the time $t$ corresponding to the flat surface 
$t=\hbox{itsvalue}$ is replaced by the general spacelike surface $\sigma$. The state 
vector is then a function $\ket{\psi(\sigma)}$ of $\sigma$. Denoting IP state vectors and 
operators by a superscript I, the Tomonaga--Schwinger evolution equation of the state 
vector is 
$$
\delta\ket{\psi^{\rm I}(\sigma)} = 
-i\,\fl H^{\rm I}_{\rm I}(x) \ket{\psi^{\rm I}(\sigma)} \delta\sigma(x)
\numera
$$
where the spacetime point $x$ belongs to $\sigma$, $\delta\ket{\psi^{\rm I}(\sigma)}$ is 
the change of the state vector in going from $\sigma$ to a nearby spacelike surface 
separated from $\sigma$ by an arbitrarily small fourdimensional bubble adjacent to $x$, 
$\delta\sigma(x)$ is the spacetime volume of the bubble, and $\fl H^{\rm I}_{\rm I}(x)$ 
is the density of IP interaction Hamiltonian at $x$. If all couplings among different 
fields are nonderivative, $\fl H^{\rm I}_{\rm I}(x)$ is automatically a Lorentz scalar. 
If derivative couplings are there, a somewhat more complicated construction is necessary 
and the role of density of interaction Hamiltonian is played by a scalar built with the 
part of the energy--momentum tensor corresponding to the interaction Lagrangian and the 
orientation of the element of $\sigma$ surrounding $x$. Eq.\ (\the\numsez.1) is then 
Lorentz invariant in all cases. 

Since one can go from a spacelike surface to another one following different paths in the 
manifold of spacelike surfaces, the evolution calculated using eq.\ (\the\numsez.1) must 
be independent of such different paths. The fulfilment of this condition of integrability 
follows from the commutativity of Hamiltonian densities at different spacelike--separated 
points. 

In a particular reference frame, one can choose the surfaces $\sigma$ to be the flat 
surfaces corresponding to arbitrary values of $t$ and investigate the change 
$\diff\ket{\psi^{\rm I}(t)}$ of the IP state vector in going from $t$ to $t+\diff t$. Let 
$C(\mbi x_i)$ be a set of disjoint arbitrarily small cubes exhausting the tridimensional 
space. If $\diff V$ is the volume of the cubes, the volume of each spacetime domain 
identified by a cube and by the time interval $(t,t+\diff t)$ is 
$$
\delta\sigma(t,\mbi x_i) = \diff V\, \diff t . 
\numera
$$
Then, according to eq.\ (4.1), 
$$
\diff\ket{\psi^{\rm I}(t)} = -i\,{\sum}_i \fl H^{\rm I}_{\rm I}(t,\mbi x_i) \diff V 
\,\ket{\psi^{\rm I}(t)} \,\diff t = 
-i \,H^{\rm I}_{\rm I}(t) \ket{\psi^{\rm I}(t)} \,\diff t , 
\numera
$$
where $H^{\rm I}_{\rm I}(t)$ is the IP interaction Hamiltonian operator in the considered 
reference frame. 

\sez{\bf Relativistic reduction equation}

In a Tomonaga--Schwinger framework, we propose the stochastic evolution equation 
$$
\delta\ket{\psi^{\rm I}(\sigma)} = \left[-i\,\fl H^{\rm I}_{\rm I}(x) \,\delta\sigma(x) 
+ g \,S^{\rm I}_{\psi(\sigma)}(x) \,\delta \beta(x)
- {\ts{1\over2}}\, g^2\!\left(\hs{-1}S^{\rm I}_{\psi(\sigma)}(x)\hs{-1}\right)^{\!2}\!
\delta\sigma(x)\right]\ket{\psi^{\rm I}(\sigma)} , 
\numera
$$
where
$$
S^{\rm I}_{\psi(\sigma)}(x) = 
S^{\rm I}(x) - \bra{\psi^{\rm I}(\sigma)} S^{\rm I}(x) \ket{\psi^{\rm I}(\sigma)}
\numera
$$
and, in correspondence to the spacetime bubble $\delta\sigma(x)$, $\delta\beta(x)$ is a 
Gaussian random variable such that 
$$
\overline{\delta\beta(x)} = 
0,\hs{40}\overline{\,\big(\hs{-1}\delta\beta(x)\hs{-1}\big)^{\!2}} = \delta\sigma(x) . 
\numera
$$
The linear operator $S^{\rm I}(x)$ is intended to be a Lorentz scalar field describing a 
macroscopic quantity associated in a proper way to the point $x$. If $S^{\rm I}(x)$ is a 
scalar, the Lorentz invariance of eq.\ (5.1) is obvious. 

When the evolution is calculated, a collection $x_i$ of spacetime points is considered 
together with the spacetime bubbles $\delta\sigma(x_i)$ adjacent to them. It is 
understood that the corresponding random variables $\delta\beta(x_i)$ are independent, 
so that 
$$
\overline{\delta\beta(x_i)}=0,\hs{40}\overline{\delta\beta(x_i)\,\delta\beta(x_j)} = 
\delta_{ij}\, \delta\sigma(x_i) . 
\numera
$$
Eq.\ (5.4) is compatible with the arbitrary smallness of the spacetime bubbles. 
In fact, given a collection of bubbles $\delta\sigma(x_i)$, let $\delta\sigma(x_{ik})$ 
be, for each $i$ and for running $k$, a decomposition in parts of $\delta\sigma(x_i)$, so 
that, in terms of volumes, 
$$
\delta\sigma(x_i) = {\sum}_k \delta\sigma(x_{ik}) . 
\numera
$$
Let $\delta\beta(x_{ik})$ be the Gaussian random variables corresponding to the finer 
collection of bubbles, such that 
$$
\overline{\delta\beta(x_{ik})}=0,\hs{40}\overline{\delta\beta(x_{ik})\, 
\delta\beta(x_{jl})} = \delta_{ij}\, \delta_{kl}\, \delta\sigma(x_{ik}) . 
\numera
$$
Then 
$$
\delta\beta(x_i) \equiv {\sum}_k \delta\beta(x_{ik}) 
\numera
$$
are Gaussian random variables which, because of eqs.\ (5.6) and (5.5) satisfy condition 
(5.4). Therefore, to each realization of the random variables $\delta\beta(x_{ik})$ there 
corresponds a realization of the random variables $\delta\beta(x_i)$ and the statistical 
distributions of such realizations agree. 

Similarly to the pure Tomonaga--Schwinger equation, the integrability of eq.\ (5.1) has 
to be demonstrated. We postpone the discussion of this problem. Assuming integrability, 
it follows from eq.\ (5.4) that the change $\delta_{\{i\}}\ket{\psi^{\rm I}(\sigma)}$ of 
the state vector in going from $\sigma$ to a nearby surface separated from $\sigma$ by a 
collection of bubbles $\delta\sigma(x_i)$ is given by 
$$\eqalign{
\delta_{\{i\}}\ket{\psi^{\rm I}(\sigma)} = 
\bigg[&-i\,{\sum}_i\fl H^{\rm I}_{\rm I}(x_i) \delta\sigma(x_i) 
\cr
&+ {\sum}_i\bigg(g \,S^{\rm I}_{\psi(\sigma)}(x_i) \,\delta \beta(x_i)
- {\ts{1\over2}}\, g^2\Big(\hs{-1}S^{\rm I}_{\psi(\sigma)}(x_i)\hs{-1}\Big)^{\!2} 
\delta\sigma(x_i)\bigg)\bigg]\ket{\psi^{\rm I}(\sigma)} . 
}
\numera
$$

In a particular reference frame, we again choose the surfaces $\sigma$ to be the flat 
surfaces corresponding to arbitrary values of $t$ and investigate the change 
$\diff\ket{\psi^{\rm I}(t)}$ of the state vector in going from $t$ to $t+\diff t$. In the 
considered reference frame, we envisage a stochastic field $B_{\mbi x}(t)$ such that 
$$
\delta\beta(t,\mbi x_i) = \!\!\int_{\!C\hs{-.5}(\!\mbi x_{\hs{-.5}i}\!)} 
\hs{-12}\diff^3\mbi x \,\,\diff B_{\mbi x}(t) , 
\numera
$$
where $C(\mbi x_i)$ are cubes as those of sect.\ 4. It is easily checked that the 
properties 
$$
\overline{\diff B_{\mbi x}(t)}=0 , ~~~~~~
\overline{\diff B_{\mbi x}(t)\,\diff B_{\mbi x'}(t)} = 
\delta^{\sss(\hs{-.5}3\hs{-.5})}\hs{-.5}(\mbi x - \mbi x\pr{1.5}) \,\diff t 
\numera
$$
imply, through relations (5.9) and (4.2), eq.\ (5.4). In fact 
$$\eqalign{
\overline{\delta\beta(t,\mbi x_i)} &= 0 , 
\cr\smalldisplayskip
\overline{\delta\beta(t,\mbi x_i)\,\delta\beta(t,\mbi x_j)} &= 
\!\!\int_{\!C\hs{-.5}(\!\mbi x_{\hs{-.5}i}\!)} \hs{-12}\diff^3\mbi x 
\,\,\!\!\int_{\!C\hs{-.5}(\!\mbi x_{\hs{-.5}j}\!)} \hs{-12}\diff^3\mbi x\pr{1.5} \,\,
\overline{\diff B_{\mbi x}(t)\,\diff B_{\mbi x'}(t)} 
\cr\smalldisplayskip
&= 
\!\!\int_{\!C\hs{-.5}(\!\mbi x_{\hs{-.5}i}\!)} \hs{-12}\diff^3\mbi x 
\,\,\!\!\int_{\!C\hs{-.5}(\!\mbi x_{\hs{-.5}j}\!)} \hs{-12}\diff^3\mbi x\pr{1.5} \,\,
\delta^3(\mbi x-\mbi x\pr{1.5})\, \diff t = \delta_{ij} \, \diff V \diff t = 
\delta_{ij} \, \delta\sigma(t,\mbi x_i) \, . 
}
\numera
$$
Then, applying eq.\ (5.8) to the calculation of $\diff\ket{\psi^{\rm I}(t)}$, 
one finds 
$$\eqalign{
\diff\ket{\psi^{\rm I}(t)} = 
\Big[&-i\,{\sum}_i\fl H^{\rm I}_{\rm I}(t,\mbi x_i) \diff V\, \diff t 
\cr
&+ {\sum}_i\Big(\!g \,S^{\rm I}_{\psi(t)}(t,\mbi x_i) \,
\!\!\int_{\!C\hs{-.5}(\!\mbi x_{\hs{-.5}i}\!)} \hs{-12}\diff^3\mbi x 
\,\,\diff B_{\mbi x}(t)- {\ts{1\over2}}\, g^2\big(\hs{-1}S^{\rm I}_{\psi(t)}(t,\mbi x_i)
\hs{-1}\big)^{\!2} \diff V\, \diff t\!\Big)\Big]\ket{\psi^{\rm I}(t)} 
\cr
=\Big[&-i\,H^{\rm I}_{\rm I}(t) \, \diff t +\compint \diff^3\mbi x 
\,\Big(\!g\,S^{\rm I}_{\psi(t)}(t,\mbi x) \,\diff B_{\mbi x}(t) - {\ts{1\over2}}\, g^2
\big(\hs{-1}S^{\rm I}_{\psi(t)}(t,\mbi x)\hs{-1}\big)^{\!2} \diff t\!\Big)
\Big]\ket{\psi^{\rm I}(t)} . 
}
\numera
$$

Eq.\ (5.12) is written in the interaction picture. We note, however, that the 
relationship between the interaction picture and the Schr\"odinger picture is much less 
trivial in the situation we are interested in than in the common applications of 
quantum field theory. In fact, in our case, bound subsystems of macroscopic systems must 
necessarily be considered and, on the other hand, the free evolution operator $U_0(t)$ 
which connects the two pictures destroys any bound state because it contains no 
interaction, so that it simply moves each constituent particle according to its momentum 
content in the bound state. As a consequence, describing bound states in the interaction 
picture is practically impossible. For this reason it is convenient to rewrite eq.\ 
(5.12) in the Schr\"odinger picture (SP). Denoting state vectors and operators in the SP 
by no superscript, the two pictures are related by 
$$
\ket{\psi(t)} = U_0(t)\, \ket{\psi^{\rm I}(t)} ,~~~~~~ 
A = U_0^{~}(t)\,A^{\rm I}(t)\,U_0^+(t)
\numera
$$
and the evolution equation for $\ket{\psi(t)}$ turns out to be 
$$
\diff\ket{\psi(t)} = \Big[-i\,H \diff t 
+\compint \diff^3\mbi x \,\Big(\!g\,S_{\psi(t)}(\mbi x) \,\diff B_{\mbi x}(t)
- {\ts{1\over2}}\, g^2 \!
\left(\hs{-1}S_{\psi(t)}(\mbi x)\hs{-1}\right)^{\!2} \!\diff t\!\Big)
\Big]\ket{\psi(t)} , 
\numera
$$
where $H$ is the total Hamiltonian operator and 
$$\leqalignno{
S_{\psi(t)}(\mbi x) &= S(\mbi x) - 
\bra{\psi(t)} S(\mbi x) \ket{\psi(t)} , 
&\global\advance\numeq by 1(\the\numsez.\the\numeq)
\cr\smalldisplayskip
S(\mbi x) &= U_0^{~}(t) \,S^{\rm I}(t,\mbi x)\, U_0^+(t) . 
&\global\advance\numeq by 1(\the\numsez.\the\numeq)
}$$

We note the formal similarity of eq.\ (5.14) to eqs.\ (3.1) or (3.9). Eq.\ (5.14) is not 
manifestly Lorentz invariant. It is, however, Lorentz invariant provided the operator 
$S^{\rm I}(x)$ appearing in eq.\ (5.16) is a Lorentz scalar. Of course, each element of 
eq.\ (5.14) must be properly transformed in going from a reference frame to another 
one. 

\sez{\bf Stuff}

The macroscopic operators $S^{\rm I}(x)$ are IP operators defined by 
$$
S^{\rm I}(x) = \!\int_{\!D(x)} \hs{-13}\diff^4 \bar x \,\sInt(\bar x) , 
\numera
$$
where $\sInt(x)$ is a Lorentz scalar field built with IP field operators at spacetime 
point $x$ and their derivatives. The integration domain $D(x)$ (Fig.\ 1) is the set of 
points $\bar x\equiv (\bar t,\bar\mbi x)$ such that 
$$
-a^2 \le (\bar t-t)^2 - (\bar\mbi x-\mbi x)^2 \le a^2,  
\numera
$$
where $a$ is a small macroscopic length which is intended to play the same role played by 
$a$ in the nonrelativistic theory of sect.\ 3. Clearly, $S^{\rm I}(x)$ is in turn a 
Lorentz scalar field. Other choices are possible for the domain $D(x)$, but we consider 
here only the definition (6.2). The microscopic operators $\sInt(x)$ will be defined 
below as representing the spatial density of a quantity we call stuff which is related to 
the presence of massive particles. Therefore the quantity represented by $S^{\rm I}(x)$ 
has the meaning of a time--integrated amount of stuff. 

The domain $D(x)$ extends to infinity in spacetime. Therefore the states 
$\ket{\psi^{\rm I}(\sigma)}$, $\ket{\psi^{\rm I}(t)}$ or $\ket{\psi(t)}$ must in 
principle be intended as states of the universe. Then the main problem with the set of 
quantities (6.1) is that they could be unable to distinguish locally different 
distributions of stuff because of the overwhelming contributions to them of remote stuff. 
We shall see in sect.\ 7 that this problem is not there.

We propose that the density of stuff is the quantum analogue of the classical quantity  
$$
\sCl(\bar x) = \Tmumu , 
\numera 
$$
i.e.\ the invariant trace of the energy--momentum tensor $T^{\mu\nu}(\bar x)$. Since 
dimensionally $\Tmumu$ is a density of energy, the time--inte\-grated amount of stuff is 
then an action. 

To illustrate the physical meaning of $\Tmumu$ and of the ensuing time--integrated amount 
of stuff we make reference to the case of classical mechanical systems. For a free 
pointlike particle moving with velocity $\mbi v$ one finds 
$$
\Tmumu = m \sqrt{1-\mbi v^2} \,
\delta^{(3)}(\bar\mbi x-\mbi x_0 -\mbi v\bar t) , 
\numera
$$
$m$ being the rest mass of the particle. The corresponding value of the time--integrated 
amount of stuff is 
$$\eqalign{
S^C\!(x) &= \!\int_{\!D(x)} \hs{-13}\diff^4 \bar x \,\sCl(\bar x)
\cr\smalldisplayskip
&= 2m \left(\sqrt{\mbi x_1^2+(\mbi x_1\ncdot\mbi v)^2/(1-\mbi v^2)+a^2}
-\sqrt{\mbi x_1^2+(\mbi x_1\ncdot\mbi v)^2/(1-\mbi v^2)-a^2}\right) , 
}\numera
$$
where $\mbi x_1=\mbi x_0+\mbi vt-\mbi x$ is the position of the particle at time $t$ 
referred to $\mbi x$ and the second square root disappears if its argument is negative. 
If the particle is in $\mbi x$ at time $t$, then $\mbi x_1=0$ and 
$$
S^C\!(x)=2ma. 
\numera
$$
If the particle is far from $\mbi x$, i.e.\ $|\mbi x_1|$ is large, we consider two 
extreme situations. First, let the trajectory of the particle be orthogonal to 
$\mbi x_1$. Then $\mbi x_1\!\ncdot\mbi v=0$ and, for large $|\mbi x_1|$, we find 
$$
S^C\!(x)=2ma\,{a\,\over|\mbi x_1|} \fracp
\numera
$$
On the opposite, if the trajectory is parallel to $\mbi x_1$, then 
$(\mbi x_1\!\ncdot\mbi v)^2=\mbi x_1^2\mbi v^2$ and, for large $|\mbi x_1|$, we get 
$$
S^C\!(x)=2ma\,\sqrt{1-\mbi v^2}\,{a\,\over|\mbi x_1|} \fracp 
\numera
$$ 
In both the extreme cases the time--integrated amount of stuff of a far particle is 
significantly reduced with respect to that of a particle in $\mbi x$ and we expect that a 
similar conclusion holds also in the intermediate situations. 

If several free particles are present, a similar contribution from each particle is 
there. In the case of a noninteracting distribution of mass at rest in the considered 
reference frame, $\Tmumu$ is the distribution of mass itself and the time--integrated 
amount of stuff is given by an integral in tridimensional space where each volume 
element is multiplied by the density of mass in that element times the extension of the 
time interval (or intervals) allowed by inequalities (6.2) in correspondence with that 
element. The farther is the element of volume from $\mbi x$, the smaller is such a time 
extension.

That we have found is just the type of behaviour we need in order that the quantities 
$S^C\!(x)$ for suitable values of $x$ be able to distiguish among different distributions 
of stuff. 

The above examples concern noninteracting systems. If interactions are there, they will 
also contribute to $\Tmumu$ and $S^C\!(x)$. 

In the case of a field the energy--momentum tensor is 
$$
T\si{6.5}{0}^{\mu\nu}= {\sum}_{\sss K} \partial^\mu\!\phi^{\sss K} 
{\partial \fl L \over \partial_\nu\hs{-.6}\phi^{\sss K}}
-g^{\mu\nu} \!\fl L + \partial_\lambda A^{\mu\nu\lambda} , 
\numera
$$
where $\fl L$ is the Lagrangian density, $\ss K$ runs over all components of the field 
$\phi$ and $\ds A^{\mu\nu\lambda}$ is an arbitrary tensor antisymmetric in the last two 
indices to be chosen so that $T\si{6.5}{0}^{\mu\nu}$ is symmetric. Except for the 
electromagnetic field, classical relativistic fields have no direct physical meaning. One 
should go to the corresponding quantum fields with their particle interpretation and, in 
such a framework, undertake the description of the world. For the electromagnetic field, 
both classical and quantum, 
$T\si{6.5}{0}^\mu\raise1.5pt\hbox{$\si{1}{0}_{\hs{-1.5}\mu}$}$ 
and consequently the time--integrated amount of stuff are identically zero. The other 
cases of interest are presently being investigated. If the meaning and behaviour of the 
time--integrated amount of stuff as it emerges from the above classical mechanical 
examples will be confirmed, we think that 
$T\si{6.5}{0}^\mu\raise1.5pt\hbox{$\si{1}{0}_{\hs{-1.5}\mu}$}$ 
is a good candidate for the density of stuff. 

\sez{\bf Stuff operators}

When interactions are present, and of course they actually are there, the quantum 
analogues of $\sCl(\bar x)$ and $S^C\!(x)$ are the corresponding Heisenberg--picture 
operators, not the interaction--picture operators $\sInt(\bar x)$ and $S^{\rm I}(x)$ 
appearing in and defined by eq.\ (6.1). The difference does not consist simply in the 
inclusion in the density of stuff $\Tmumu$ of the contributions coming from the 
interactions, but lies in the fact that defining the time--integrated amount--of--stuff 
operator in one picture of time evolution or another is not the same thing. This 
inequivalence is a consequence of the time integration present in  the definition (6.1) 
and such a time integration is necessary if $S^{\rm I}(x)$ has to be a Lorentz scalar. 

Choosing a reference frame and identifying $\sigma$ with time $t$, let us write down 
explicitly the Schr\"odinger--picture operator $S(\mbi x)$ related to $S^{\rm I}(x)$ by 
eq.\ (5.16). From 
$$
S^{\rm I}(t,\mbi x) = \!\int_{-\infty}^{+\infty} \hs{-13}\diff \bar t
\int_\domscript \hs{-19}\diff^3 \bar\mbi x \,\sInt(\bar t,\bar\mbi x) , 
\numera
$$
where {\domtext} is the sphere or the spherical shell of points $\bar \mbi x$ satisfying 
condition (6.2) for fixed $\bar t$, denoting again by no superscript SP operators one 
gets  
$$\eqalign{
S(\mbi x) &= U_0^{~}(t) \!\int_{-\infty}^{+\infty} \hs{-13}\diff \bar t
\int_\domscript \hs{-20}\diff^3 \bar\mbi x \, U_0^{+}(\bar t\hs1)
\,s(\bar\mbi x) \, U_0^{~}(\bar t\hs1) \,U_0^+(t) 
\cr \smalldisplayskip 
&= \!\int_{-\infty}^{+\infty} \hs{-13}\diff \bar t \,\,U_0^+(\bar t-t) \!\!
\int_\domscript \hs{-20}\diff^3 \bar\mbi x \,s(\bar\mbi x) \,U_0^{~}(\bar t-t) . 
}\numera
$$
It is easily checked that $S(\mbi x)$ is actually independent of $t$. If we should have 
written eq.\ (6.1) in the Heisenberg picture, the full time--evolution operator $U$ would 
appear in eq.\ (7.2) instead of 
$U_0$. 

It is seen that when the operator $S(\mbi x)$ is applied to the SP state $\ket{\psi(t)}$ 
one must evolve freely the state from time $t$ to time $\bar t$, apply the SP operator 
having the meaning of amount of stuff contained in the domain {\domtext} and then evolve 
back the result, again freely, up to the time $t$. Finally one integrates over time 
$\bar t$. 

We must now understand whether the set of operators $S(\mbi x)$ does the job we want it 
do. 

We first consider a local system, i.e.\ a system contained in a finite spatial domain, 
the system's environment being forgotten. Let the SP state at time $t$ be 
$$
\ket{\psi(t)} = {\sum}_\alpha\, c_\alpha \ket{\lambda_\alpha(t)} , 
\numera
$$
where the states $\ket{\lambda_\alpha(t)}$ are macroscopically distinguishable on the 
basis of their different macroscopic distributions of stuff. 

To allow for a simple description of what is going on, we assume that the system consists 
of a single macroscopic object. Each term in expansion (7.3) describes a unique bound 
system whose center of mass is in a definite state different for different terms. When 
the free time--evolution operator $U_0(\bar t-t)$ is applied, this structure is 
maintained in the sense that the center of mass remains in a definite state and moves 
accordingly. However, the object as a bound state is shattered by $U_0$, as we have 
already discussed in sect.\ 5. In order to estimate the result of the application of the 
operator (7.2) to the different terms in expansion (7.3), we suppose that the envisaged 
object is at rest in the considered reference frame and in two different positions for, 
say, $\alpha=1$ and $\alpha=2$ (Fig.\ 2). Some amount of stuff $\cal S$ belonging to the 
object lies at $\mbi x$ for $\alpha=1$ and at a distance $l$ from 
$\mbi x$ for $\alpha=2$. Let the amount of stuff $\cal S$ be projected around 
isotropically with velocity $w$ by the operator $U_0(\bar t-t)$. Then the 
time--integrated amount of stuff counted by the operator (7.2) is easily estimated and 
turns out to be 
$$
\matrix{&~~~~2a{\cal S} \,\left[\ds{{1\over\sqrt{1-w^2}}}\right],~~~~ 
&~~~~2a{\cal S}\,\ds{{a\over l}} \left[{2\over\pi}\,K\!\left(w^2\right)\right],~~~~
\cr\displayskip
&(\alpha=1) &(\alpha=2)}
\numera
$$
where the expression for the case $\alpha=2$ is evaluated for $l^2\gg a^2$ and the 
function $K(\,\cdot\,)$ is the complete elliptic integral of the first kind. The factor 
in square brackets is always smaller for $\alpha=2$ than for $\alpha=1$. Therefore, in 
the case of the considered example, the set of operators (7.2) discriminates the various 
terms in expansion (7.3). If the object is moving, and $\cal S$ with it, the above 
argument is to be supplemented by the discussion in sect.\ 6, the particle considered 
there being identified with the center of mass of $\cal S$. We think it is evident that 
our conclusion remains true for general superpositions of macroscopically distinguishable 
states. 

As already remarked, the states in our equations are in principle states of the universe, 
so that, besides the local system, we must take into account the rest of the universe, 
which we call the environment. For large $|\bar t-t|$ the domain {\domtext} is a 
spherical shell centered in $\mbi x$ whose volume is $4\pi|\bar t-t|a^2$. The first 
problem which then arises concerns the convergence of the integrals in the definition of 
$S(\mbi x)$. This problem disappears if we assume that the universe is finite, but 
another problem remains. If we include the environment in the description, its 
contribution to the value of $S(\mbi x)$ will be overwhelmingly larger than the 
contribution of the local system, so that the values of $S(\mbi x)$ corresponding to 
macroscopically distinguishable states of the local system will no more differ 
significantly. As a consequence the stochastic process in eq.\ (5.14) could become 
ineffective. 

To discuss the problem presented above, we assume that the splitting between local system 
and environment is such that the particles in the environment can be considered as 
distinguishable from the particles in the local system even when they are of the same 
kind. This does not mean that the particles in the two subsystems of the universe do not 
interact, it means that the dynamics is such that they never come together in space. We 
do not think that this assumption is essential, but it considerably simplifies the 
discussion which follows. If particles of the same kind in the two subsystems are 
distinguishable, they can be described by distinct fields and, as a consequence, the SP 
state of the universe can be written as 
$$
\ket{\psi(t)} = {\sum}_\alpha\, c_\alpha\, \ket{\lambda_\alpha(t)} 
\,\ket{\epsjlon_\alpha(t)} , 
\numera
$$
where $\ket{\lambda_\alpha(t)}$ and $\ket{\epsjlon_\alpha(t)}$ indicate states of the 
local system and of the environment, respectively. The free evolution operator $U_0(t)$ 
is the product of an operator acting on the states $\ket{\lambda_\alpha(t)}$ and an 
operator acting on the states $\ket{\epsjlon_\alpha(t)}$. May be that some particles of 
the environment and some particles of the local system are brought to overlap by the free evolution operator 
$U_0(\bar t-t)$. The SP microscopic density--of--stuff operator $s(\bar\mbi x)$ will 
contain terms coming from the free Lagrangian density of the fields and terms coming from 
the interaction. If the contribution of the interaction between local--system fields and 
environment fields is disregarded in $s(\bar\mbi x)$, then the SP time--integrated 
amount--of--stuff operator is the sum 
$$
S(\mbi x) = L(\mbi x) + E(\mbi x) , 
\numera
$$
of a term $L(\mbi x)$ acting on the local states $\ket{\lambda_\alpha(t)}$ and a term 
$E(\mbi x)$ acting on the environment states $\ket{\epsjlon_\alpha(t)}$. In the situation 
in which we are interested, the local states $\ket{\lambda_\alpha(t)}$ are 
distinguishable on the basis of their different macroscopic distributions of stuff. The 
environment states $\ket{\epsjlon_\alpha(t)}$ are mutually orthogonal if decoherence is 
there, as it is reasonable to assume. Nevertheless, they are macroscopically 
undistinguishable, so that 
$$
E(\mbi x) \,\ket{\epsjlon_\alpha(t)} = 
e(t,\mbi x) \,\ket{\epsjlon_\alpha(t)} , 
\numera
$$
with the eigenvalues $e(t,\mbi x)$ independent of $\alpha$. It is then easily shown that 
the two contributions of $E(\mbi x)$ in eq.\ (5.15) cancel each other, so that 
$$\eqalign{
\Big(S(\mbi x) - \bra{\psi(t)}S(\mbi x)\ket{\psi(t)}\Big)\ket{\psi(t)} &= 
\left(L(\mbi x) - {\sum}_\alpha |c_\alpha|^2 
\bra{\lambda_\alpha(t)}L(\mbi x)\ket{\lambda_\alpha(t)} \right) \ket{\psi(t)} 
\cr\smalldisplayskip
&= \Big(L(\mbi x) - \bra{\psi(t)}L(\mbi x)\ket{\psi(t)}\Big)\ket{\psi(t)} , 
}
\numera
$$
and similarly when the bracket is squared. 

We conclude that the set of operators $S(\mbi x)$ is able to discriminate among states 
having locally different macroscopic distributions of stuff even when the states are 
states of the universe and, as a consequence, the stochastic term in eq.\ (5.14) causes 
the type of reduction we are interested in. 

\sez{\bf Open problems and conclusion}

So far we took for granted that the time--integrated amount--of--stuff operators (6.1) 
describe classical macroscopic quantities. This means that they should mutually commute 
to a good approximation and that they should not exhibit appreciable fluctuations related 
to the microscopic structure of the system. This behaviour should be a consequence of the 
choice of the constant $a$, defining the integration domain $D(x)$, as a small but 
macroscopic length. We have few doubts that an operator like the field--theoretic 
$\Tmumu$ integrated over a sufficiently large spacetime domain possesses such properties. 
However, the space domain {\domtext} corresponding for given $\bar t$ to the spacetime 
domain $D(x)$ is, for large $|\bar t-t|$, a spherical shell of radius $|\bar t-t|$ and of 
vanishingly small thickness $a^2/|\bar t-t|$. We hope to be able to prove that this 
particular shape of the integration domain does not spoil the classical macroscopic 
features of the operators (6.1). We are presently working on this problem. 

As mentioned in sect.\ 5, the result of the application of the modified 
Tomonaga--Schwinger equation (5.1) should be independent of the path in the manifold of 
spacelike surfaces followed in going from a spacelike surface to another one. We do not 
know whether there exists a general mathematical theory allowing to deal with such an 
integrability problem. In the case of the pure Tomonaga--Schwinger equation, the 
commutativity of the operators $\fl H^{\rm I}_{\rm I}(x)$ at different spacelike 
separated points is likely to be sufficient to ensure integrability.  Similarly, if the 
Schr\"odinger term is dropped from eq.\ (5.1), the commutativity of operators 
$S^{\rm I}(x)$ related to their classical nature should be sufficient for integrability. 
If both the Schr\"odinger and the stochastic term are retained, however, the 
commutativity argument breaks down, because, considered $S^{\rm I}(x)$ and 
$\fl H^{\rm I}_{\rm I}(x\pr{1.5})$, $S^{\rm I}(x)$ contains contributions from points 
which are not spacelike separated from the point $x\pr{1.5}$ even if $x$ and $x\pr{1.5}$ 
are spacelike separated. We note, however, that the stochastic process is effective only 
when superpositions of macroscopically distinguishable states are there and that, when 
such a kind of superposition starts to be created in the region around $x\pr{1.5}$, 
$S^{\rm I}(x)$ is not able (is little able) to distinguish among the superposed terms if 
$x$ is away from $x\pr{1.5}$. Therefore, advancing the spacelike surface first around 
$x\pr{1.5}$ and then around $x$ or viceversa should be unimportant. We plan to discuss 
this point in model situations. 

The stochastic equation (5.1) and the time--integrated amount--of--stuff operators (6.1) 
contain altogether two new constants, the length $a$ and the strength constant $g$. As 
already said, $a$ has to be a small macroscopic length. The same value $10^{-5} {\rm cm}$ 
proposed for the nonrelativistic model is a reasonable choice. Suggesting a value for $g$ 
is much more difficult, even though one could try to get a hint from the value of the 
strength constant for the nonrelativistic model, using the formal similarity of 
eq.\ (5.14) to eq.\ (3.9). Of course, the goal is again getting from the stochastic 
process a rapid suppression of superpositions of macroscopically distinguishable states 
and negligible effects for all the rest. The evaluation of the physical consequences of 
our model, depending on the value of $g$ is a job to be done. 

Finally, the interpretative implications of the type of theory we are proposing are to be 
investigated. 

Concluding, we do not presently have definite and definitive answers to the problems 
presented above. In any case we are convinced that the right road to build a relativistic 
reduction theory is that of identifying a suitable set of macroscopic quantities to be 
stochastically compelled to have definite values, because this is, we think, the physical 
meaning of reduction. 

\vs{18} 
\leftline{\bf Acknoledgement} 
\vs9 

We warmly thank Guido Montagna for invaluable discussions and suggestions. 

\vs{18}
\leftline{\bf References}
\vs9
\immediate\closeout\notefile%
\parindent=20pt%
\frenchspacing\input notes\nonfrenchspacing\parindent=10pt

\vfill\eject 

\eightpoint\parindent5pt 

\centerline{\epsfbox{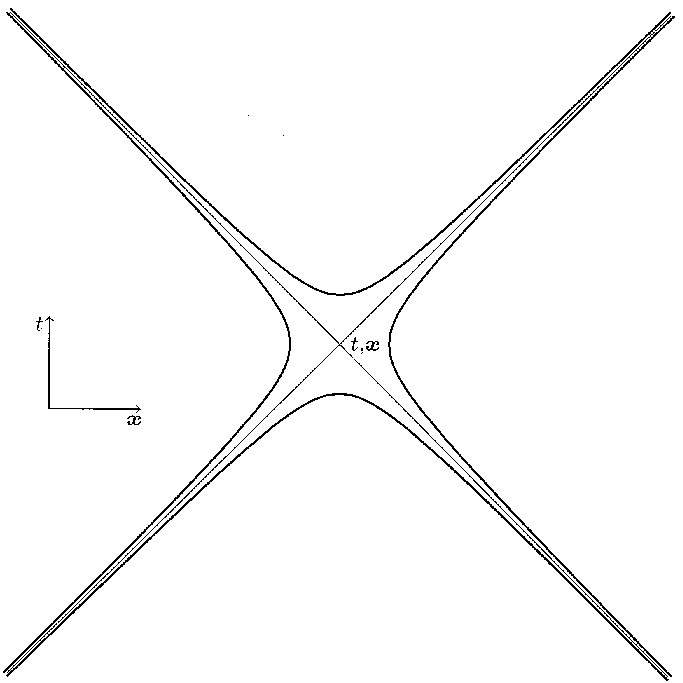}}
\vs{20}
\centercol{8}{
Fig.\ 1. The integration domain $D(x)$ represented in a bidimensional spacetime. 
}

\vs{60}

\centerline{\epsfbox{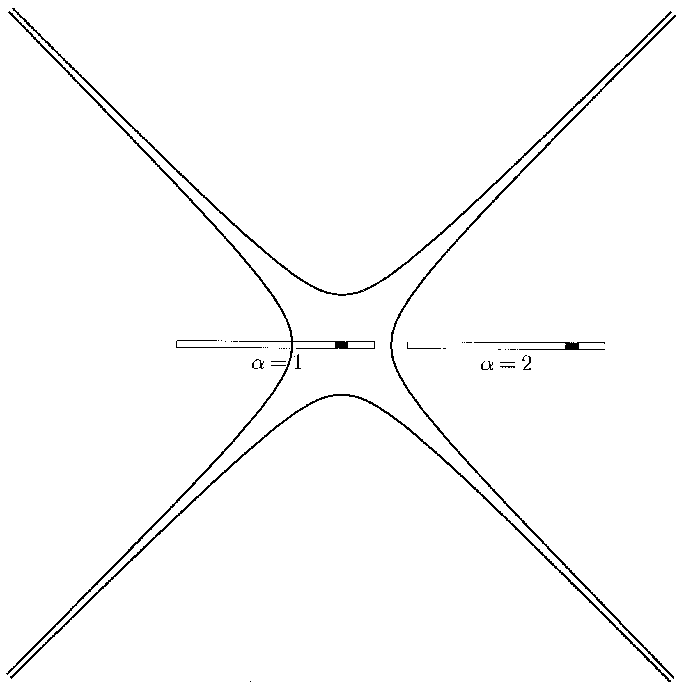}}
\vs{20}
\centercol{8}{
Fig.\ 2. A macroscopic object in two different positions. The thick dash represents an 
amount of stuff belonging to the object.} 

\tenpoint 

\bye